# Development and Performance of a Static Pluviometer System

Parth Saxena, Pratham Saxena, Adarsh Sowcar, Sreeharsha Angara, Ashutosh Pandey, Thomas Basikolo

**Abstract:** As the frequency and severity of climate-related events such as droughts, floods, and water scarcity continue to escalate, accurate rainfall monitoring becomes increasingly critical. This paper covers various industry methods of measuring rainfall as well as our own ground pluviometer system. Our system consists of an inexpensive static rain gauge that can operate for approximately six to twelve months without maintenance. It utilizes resistive sensing technology accompanied by a microcontroller to measure the water level depth from the device vessel, recording rainfall at an hourly rate. This study also provides a side-by-side comparison of our pluviometer system with an industry rain gauge, the MeteoRain 200 Compact, from Barani Systems, with the differences in data being statistically insignificant. By prioritizing cost, sustainability, simplicity, ease of maintenance, and assembly, this research contributes to essential rainfall monitoring solutions, specifically for developing countries.

## INTRODUCTION

Beyond safeguarding against these environmental risks, precise rainfall monitoring is important for agricultural practices, urban planning decisions, water resource management strategies, and disaster preparedness measures.[1-3] As the risks of droughts, floods, and scarcity of water increase, the need for accurate rainfall monitoring becomes crucial.[1,4] Developing nations often don't have access to accurate rainfall mapping due to a lack of infrastructure and a surplus of areas with poor rain gauge or weather radar coverage.[2,5] A prerequisite to encountering these challenges is the availability of rainfall data by robust ground-based pluviometer systems.[5,6,7]

For urban planning decisions, precise rainfall data is critical for flood mitigation, helping design effective drainage systems to prevent urban flooding and informing the construction of resilient infrastructure such as roads, bridges, and buildings.[4,5,6] In agricultural practices, it assists with crop planning by helping farmers determine the best times for planting and harvesting, thereby optimizing yields and minimizing losses.[2,8] In terms of disaster preparedness measures, accurate real-time rainfall data is vital for developing early warning systems for floods, typhoons, and landslides, giving communities the necessary time to evacuate and prepare.[5,7,9] Poor rain gauge or weather radar coverage leads to gaps in data, hindering effective decision-making.[5] Implementing robust ground-based pluviometer systems can bridge this gap, providing reliable data.[1,4,6]

In this paper, we present our automatic pluviometer system consisting of a low-cost, no moving parts, maintainable, and accurate rain gauge. We compare its performance to an industry-standard rain gauge, the MeteoRain 200 Compact, developed by Barani Systems. Our rain gauge is suited for areas with heavy to extreme rainfall and is to be maintained once every six months to a year. The device is modular and the rainfall depth data (in millimeters of rain) is transmitted via Arduino Uno every hour. We aim to couple such devices with machine learning models to develop smart rain gauges that can contribute to automated, controlled, and data-driven irrigation to increase crop yield and conserve water in farms.

## EXISTING SOLUTIONS

Although there are several existing high-performant pluviometer systems, these systems are often composed of multiple complex components in a dynamic state. Two widely accepted rain gauges in this category are the Weighing Rain Gauge and the Tipping Bucket Rain Gauge. As shown in Figure 1, the Weighing Rain Gauge measures rainfall by collecting the precipitated water onto a surface (typically a bucket) and detecting the measured weight relative to time, thus deriving the depth of the rainfall.[10] However, this system requires high energy cost and maintenance (emptied by staff), making it unsuitable for agricultural usage or installation in remote locations.[11] The Tipping Bucket Rain Gauge is an apparatus that releases water once a bucket with a known volume is full, by tipping over, and it determines rainfall from the number of times it has tipped over by using a magnetic sensor to detect each tipping event and send an electrical signal to record the total number of tips.[10] This rain gauge is demonstrated in Figure 2, and the Barani MeteoRain 200 Compact, the device we used for comparison testing, utilizes this design and has a minimum resolution of 0.2 mm. Other solutions include Acoustic Precipitation measuring, which relies on the sound of droplets striking a metal surface. Optical Rain Sensor rain gauges also exist, which depend on visual cues to calculate the rainfall depth, but have recorded errors of up to 37%.[11]

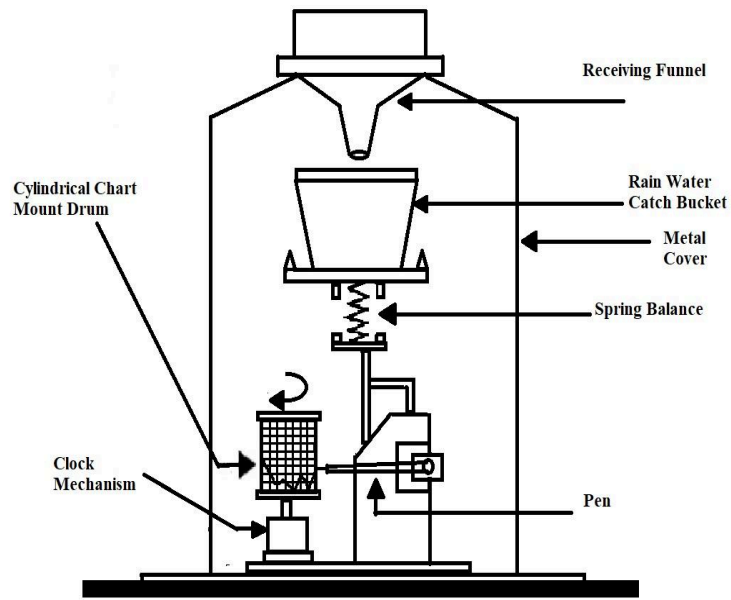

*Figure 1. A diagram of a Weighing Precipitation Sensor.*

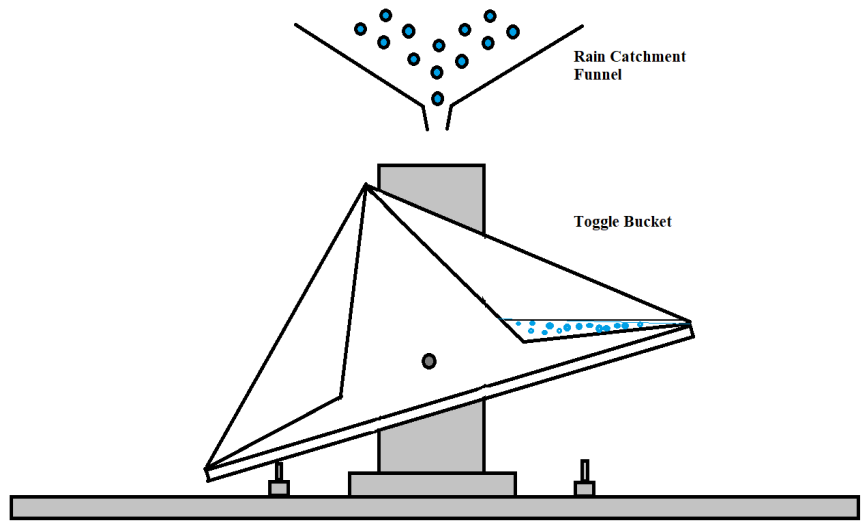

*Figure 2. A diagram of a Tipping Bucket rain gauge*

## SYSTEM COMPONENTS

This section introduces the general purpose and design considerations of the pluviometer system, and discusses its specific components and functionality, including how these elements contribute to the overall performance and reliability.

Collection Vessel

The most popular catchment area for professional use of a standard rain gauge is 200 cm², as it is the minimum acceptable size while providing reasonable accuracy. Smaller catchment areas typically result in finer resolutions and higher precision with low amounts of rain. Due to their sensitivity, high-resolution rain gauges are susceptible to errors caused by particles of dirt and debris, thus requiring constant maintenance. The largest standard catchment area is 1000 cm² and has a radius of 17.8 cm. Larger catchment areas result in low-resolution rain gauges, best suited for areas with high rainfall intensity. Conversely, rain gauges with smaller catchment areas are best suited for areas with low rainfall intensity. We decided to use larger measurements for the collection and measurement vessels to make the apparatus larger and contain higher volumes of water. This is due to the assumption that in underdeveloped/remote locations, the maintenance frequency would be low, roughly six to twelve months.

## Measurement Vessel

The measurement vessel in pluviometer systems is another major element. The size of the measurement vessel affects the amount of precipitated water remaining in the vessel after rainfall. As shown in Figure 3, our measurement vessel is cylindrically shaped with a height of 89 cm and a radius of 7.9 cm. The reasoning for the size of the measurement vessel was similar to the collection vessel: to be capable of measuring and storing large amounts of water with low maintenance.

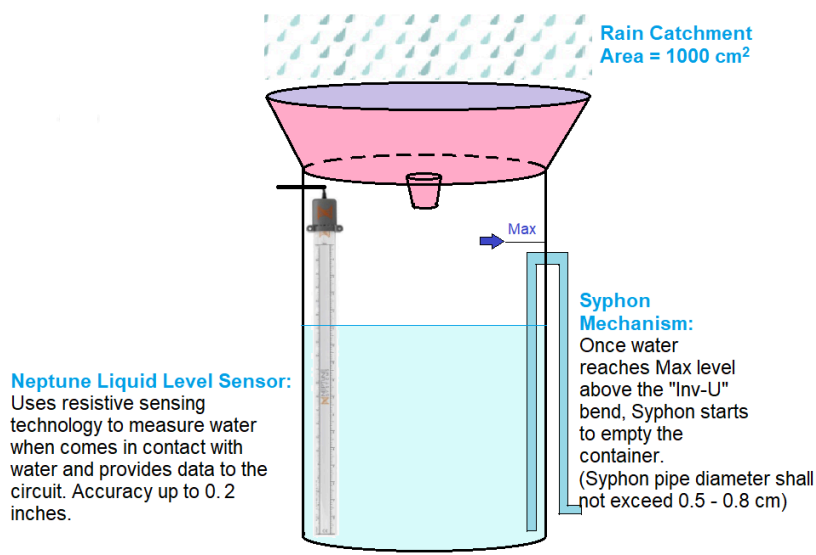

*Figure 3. Sketch of Apparatus.*

## Liquid Level Sensor

We utilized the Liquid Level Sensor (LLS) from Neptune Systems at 89 cm length to measure the water level of the Measurement Vessel to then eventually determine the millimeters of rain in the past hour. We reverse-engineered the depth sensor to correlate water depth to resistance. The LLS uses resistive sensing technology to provide water depth at a resolution of 0.51 cm. The LLS uses a TRRS cable to connect to the A2 or A3 Neptune device and transmit the data, which is then displayed on the Apex Fusion app. However, we did not use the A2/A3 Neptune device or the Apex Fusion app.

### Resolution

When testing the overall device and collecting data from the Rain Meter, we determined a minimum rain resolution of 1 mm. To calculate and reach this conclusion, we utilized the equation below:

$$Rain\ Resolution\ =\ \frac{Measurement\ Vessel\ Volume}{Collection\ Vessel\ Area}$$

As mentioned in the Collection Vessel section, the catchment area is 1000 cm², with a radius of 17.8 cm. Meanwhile, to determine the volume of the measurement vessel, we utilized the formula below to solve for the volume of a cylinder. Given a radius of 7.9 cm and a minimum height resolution of 0.51 cm, the measurement vessel has a volume of 99.99 cm³.

$$Volume\ =\ \pi\ *\ (7.9\ cm)^2\ *\ (0.51\ cm)\ =\ 99.99\ cm^3$$

With the Measurement Vessel's resolution determined, we divided it by the catchment area to retrieve the rain gauge's resolution value.

$$Rain\ Resolution\ =\ \frac{99.99\ cm^3}{1000\ cm^2}\ =\ 0.099\ cm\ \approx\ 1\ mm$$

This resolution of 1 mm makes this device less precise than other state-of-the-art rain gauges, putting it at a typically lower-resolution. However, lower-resolution rain gauges with larger catchment areas require less frequent maintenance and are more robust against environmental factors like wind and debris, making them suitable for remote locations with limited maintenance capabilities. Additionally, they offer reliability in areas prone to high rainfall intensity, making them a practical choice for measuring rainfall in such environments.

## Ohm-Meter

After conducting several tests with a multimeter, we recognized an inversely proportional linear relationship between the resistance output of the TRRS cable and the water depth.

$$Water\ Depth\ =\ -0.004(Resistance)\ +\ 21.4$$

This relationship is demonstrated below in Figure 4.

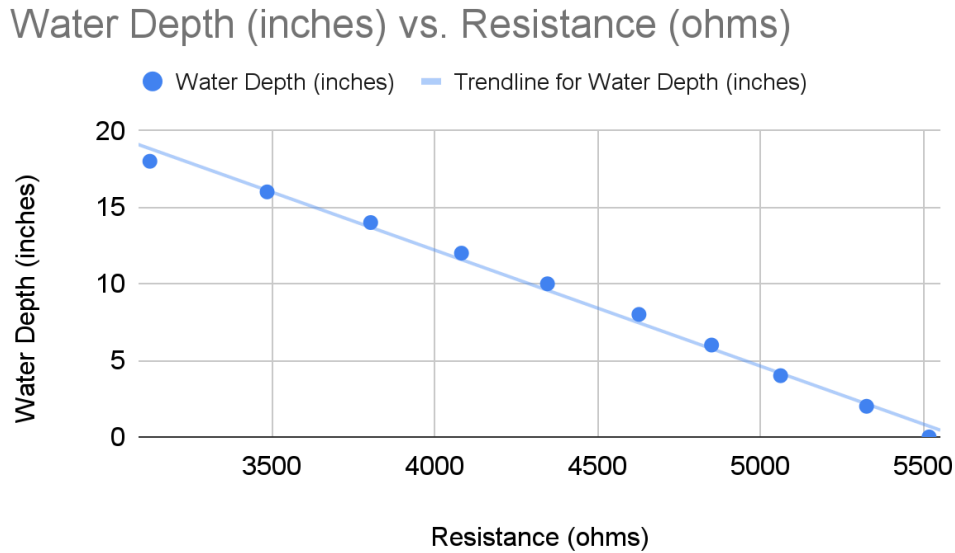

*Fig. 4. Resistance relationship to water depth*

We reduced costs by implementing an Ohm-meter module to measure resistance, translating that to water depth. As shown in Figure 5, we utilized an Arduino Uno, a known resistor of 4700 Ohms, and the unknown resistance being outputted by the LLS. We then set up a voltage divider with the known and unknown resistors, and measured the voltage between them with the Arduino, calculating the resistance from Ohm's Law. We utilized a 4700 Ohm known resistor because the accuracy of the Ohm meter would diminish if the value of the known resistor deviated significantly from the resistance output of the LLS.

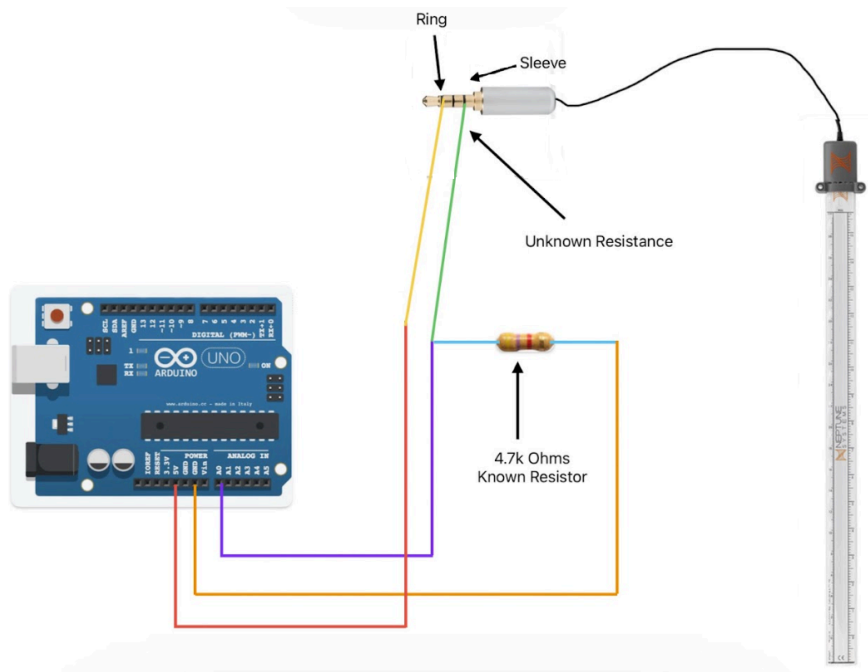

*Figure 5. Arduino Uno Resistivity Measurement Ohm-meter Setup for the depth sensor.*

## SOFTWARE

Our software was written in C, onto the Arduino Uno, and runs in a loop when the Arduino is powered on or reset. It initializes serial communication, allowing data to be interfaced with a computer. It reads an analog value from pin 0 and calculates the output voltage and the unknown resistor/LLS resistance value. These values are then printed to the Serial Monitor every minute using the line of best-fit equation established in Figure 4, and the water depth is then interpolated from the resistance value. This water depth is then translated to the volume of water held in the measurement vessel and the value of the volume is stored. After using a sliding difference, the measurement vessel's change in volume in the past hour is retrieved, which is then translated to millimeters of rain in the past hour.

## TESTS & RESULTS

For a testing reference, we utilized the Barani Meteo Rain 200 Compact device, an industry-standard device for accurate rain measurement.[11] The device operates on a tipping bucket system and has a minimum resolution of 0.2 mm of rainfall.

The testing procedure was quite simple: we placed both devices outside in San Diego, California, during days of precipitation. Below is the data collected from both devices over a 12-hour period:

| Time | MeteoRain 200 Compact hourly rain (mm) | Tested Rain Gauge hourly rain (mm) |
|:---:|:---:|:---:|
| 21:45 | 0.041 | 0.000 |
| 22:45 | 0.12 | 0.11 |
| 23:45 | 0.23 | 0.25 |
| 24:45 | 0.12 | 0.097 |
| 1:45 | 0.1 | 0.11 |
| 2:45 | 0.14 | 0.18 |
| 3:45 | 0.12 | 0.097 |
| 4:45 | 0.1 | 0.097 |
| 5:45 | 0.061 | 0.084 |
| 6:45 | 0.02 | 0.000 |
| 7:45 | 0.000 | 0.000 |
| 8:45 | 0.000 | 0.000 |

*Table 1. Data observations from Barani MeteoRain 200 Compact device and tested system*

To assess the performance of our device compared to the Barani rain gauge, we utilized a two-tailed t-test, as it is used to assess whether the mean of the expected values significantly differs from the mean of the observed values, and to determine if the means are equal, greater, or lesser. The two-tailed t-test allowed us to evaluate our hypotheses as to whether there were statistically significant differences between the hourly rain measurements of the two devices.

Hypotheses:
- Null Hypothesis ($H_0$): There is no significant difference between the Barani Meteo Rain 200 Compact's hourly rain measurements and our rain gauge's hourly rain measurements.

9- Alternative Hypothesis ($H_1$):: There is a significant difference between the Barani Meteo Rain 200 Compact's hourly rain measurements and our rain gauge's hourly rain measurements.

Since the two groups of data are independent samples of each other, the degrees of freedom are calculated as

$$df = n_1 + n_2 - 2$$

$n_1$ = number of observations from MeteoRain 200 Compact rain gauge

$n_2$ = number of observations from tested rain gauge

There are 12 categories for each hour for both groups measured, therefore $n_1 = 12$, $n_2 = 12$, and degrees of freedom are:

$$df = 12 + 12 - 2 = 22$$

We utilized a significance level, α, of 0.05. Thus, according to the two-tailed critical values distribution chart, the critical value is 2.074:

$$\alpha = 0.05$$

$$t_{critical} = 2.074$$

The T-test statistic formula is as follows:

$$t = \frac{\overline{x_1} - \overline{x_2}}{\sqrt{\frac{s_1^2}{n_1} + \frac{s_2^2}{n_2}}}$$

$\overline{x_1}$ = Observed mean of MeteoRain 200 Compact group

$\overline{x_2}$ = Observed mean of tested rain gauge group

$s_1$ = Standard Deviation of MeteoRain 200 Compact group

$s_2$ = Standard Deviation of tested rain gauge group

$n_1$ = number of observations from MeteoRain 200 Compact rain gauge

$$n_2 = number\ of\ observations\ from\ tested\ rain\ gauge$$

If the t-value is less than the critical value of 2.074, the result is within the range of typical values, and we fail to reject the null hypothesis. This means that the observed sample mean is not significantly different from the hypothesized population mean, given the sample data and the level of significance. Therefore, if the t-value is greater than the critical value (in absolute terms), the result is in the extreme tails of the t-distribution, and we reject the null hypothesis. This indicates that the observed sample mean is significantly different from the hypothesized population mean.

The calculated variables required to solve for the t-value are below:

$$t = \frac{0.089 - 0.084}{\sqrt{\frac{0.068^2}{12} + \frac{0.077^2}{12}}} \simeq 0.15$$

$$\overline{x_1} \simeq 0.089$$

$$\overline{x_2} \simeq 0.084$$

$$s_1 \simeq 0.068$$

$$s_2 \simeq 0.077$$

$$n_1 \simeq 12$$

$$n_2 \simeq 12$$

Therefore, our calculated t-value is roughly 0.15, which is significantly less than the critical value:

$$t < t_{critical}$$

The two-tailed t-test shows that the difference in hourly rain measurements between the Barani Meteo Rain 200 Compact and the tested rain gauge is not statistically significant. This implies that the evidence from the sample is not strong enough to conclude a substantial difference between the two devices.

**SIPHON ADDITION**

The final iteration of our rain gauge can be seen in Figure 6. We decided to use a siphon to automate emptying the measurement vessel once full as it does not use any moving parts and therefore wouldn't be susceptible to dust buildup or rust interference. It follows a Pythagorean Cup approach and once the water level approaches the very top (crosses the siphon pipe bend), it initiates the siphon flow and release of water until the vessel is empty.[12] The siphoning effect causes the cup to drain its entire contents through the base. At the base, water is released in a laminar flow which ensures the smooth and controlled movement of liquid through the siphon mechanism that makes the cup function. The siphon mechanism relies on a continuous, steady flow of liquid to create the necessary pressure difference that drives the liquid up, which is why the size of the tube is important to maintain the state of laminar flow.

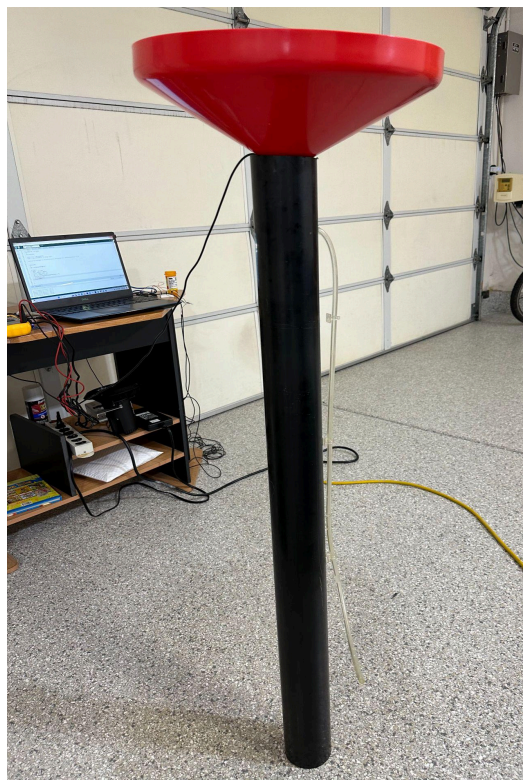

*Figure 6. Final iteration of the apparatus with siphoning mechanism.*

This allowed us to reduce the cross-section area of the measurement vessel and increase the accuracy of the rain gauge since a lower volume of water would result in increased change in water depth and result in more precise measurements of volume from the LLS. The radius of the apparatus with the siphon device was 3.25 cm, resulting in a more precise resolution of 0.169 mm, below the standard minimum rain gauge resolution set by the National Weather Station of 0.25 mm of rain, as well as the the resolution of 0.2 mm of the Barani MeteoRain 200 Compact device.[11] This was done while preserving the low maintenance of the previous iteration.

## DISCUSSION & CONCLUSION

As demonstrated by the benchmark testing with the Barani MeteoRain 200 Compact device, our rain gauge outputs rainfall data that is statistically insignificant to the industry-standard device. With the siphon technology implementation, we have further improved the device by increasing accuracy and reducing the large measurements by reducing the measurement vessel area, thus increasing the minimum resolution of the device and still having low maintenance. Our device provides continuous measurements while other approaches, like the tipping-bucket rain gauge, provide discrete measurements, since if the bucket has not tipped, rainfall is not recorded. Furthermore, due to a lack of moving parts, our rain gauge is less susceptible to dust buildup, rust, and unexpected parts interfering with accurate data output.

To create a centralized system of such rain gauges, we propose utilizing LoRa technology. LoRA technology, known for its long-range communication capabilities and low power consumption, is ideal for transmitting data from multiple rain gauges to a central hub. By integrating LoRA, each rain gauge can send hourly rainfall data over long distances without frequent battery replacements. This centralized system facilitates comprehensive rainfall mapping and ensures that data from remote locations is consistently and reliably collected. LoRA thus provides a robust framework for a network of smart rain gauges that are both energy-efficient and capable of operating autonomously for extended periods.

In conclusion, this pluviometer system addresses crucial environmental monitoring needs, offering a sustainable solution for agricultural practices, urban planning, water resource management, and disaster preparedness in developing nations. Our work in developing a sustainable and scalable solution to the critical need for accurate rainfall monitoring addresses both environmental challenges and practical constraints faced by developing nations.